# Thermodynamic and transport properties of semiconducting two-dimensional metal-organic kagomé lattices with disorder

Tanya Berry[a,b], Jennifer R. Morey[a,b], Kathryn E. Arpino[c], Jin-Hu Dou[d], Claudia Felser[c], Mircea Dincă[d], Tyrel M. McQueen[a,b,e]

- [a.] Department of Chemistry, Johns Hopkins University, Baltimore, Maryland 21218, United States
- [b.] Institute for Quantum Matter and the Department of Physics and Astronomy, Johns Hopkins University, Baltimore, Maryland 21218, United States.
- [c.] Max Planck Institute for Chemical Physics of Solids, 01187 Dresden, Germany.
- [d.] Department of Chemistry, Massachusetts Institute of Technology, Cambridge, Massachusetts 02139, USA.
- [e.] Department of Materials Science and Engineering, Johns Hopkins University, Baltimore, Maryland 21218, United States.

**Abstract:** The kagomé lattice is a fruitful source of novel physical states of matter, including the quantum spin liquid and Dirac fermions. Here we report a structural, thermodynamic, and transport study of the two-dimensional kagomé metal-organic frameworks $Ni_3(HIB)_2$ and $Cu_3(HIB)_2$ (HIB = hexaiminobenzene). Magnetization measurements yield Curie constants of 1.12 and 0.352 emu K mol f.u.$^{-1}$ Oe$^{-1}$ respectively, close to the values expected for ideal S=1 and S=1/2 moments. Weiss temperatures of -20.3 K and -6.52 K, respectively, indicate moderate to weak magnetic interactions. Electrical transport measurements reveal that both materials are semiconducting, with gaps of $E_g$ = 22.2 and 103 meV, respectively. Specific heat measurements reveal a large T-linear contribution of γ = 148(4) mJ mol-f.u.$^{-1}$ K$^{-2}$ in $Ni_3(HIB)_2$ with only a gradual upturn below T ~ 5 K and no evidence of a phase transition to an ordered state down to T = 0.1 K. $Cu_3(HIB)_2$ also lacks evidence of a phase transition above T = 0.1 K, with a substantial, field-dependent, magnetic contribution below T ~ 5 K. Despite being superficially in agreement with expectations of magnetic frustration and spin liquid physics, we are able to explain these observations as arising due to known stacking disorder in these materials. Our results further state the art of kagomé lattice physics, especially in the rarely explored regime of semiconducting but not metallic behavior.

**Introduction:**

The kagomé lattice, consisting of corner-sharing triangles of atoms, is known in the metallic case to host Dirac fermions or electrons with a linear energy-momentum relationship. These states are often topological and known to give rise to a giant anomalous Hall effect,[1-3] and, more recently, superconductivity.[4-7] In the highly insulating limit, kagomé lattices have long been explored as candidates for hosting the quantum spin liquid state.[8-10]

Metal-organic frameworks (MOFs) are a class of materials defined by metal centers connected by organic linkers. The nature of the metal and the organic linkers and the structure dictate the material's properties. Long explored for gas sorption properties, MOFs have recently received appreciable attention due to graphene-like structural motifs and high carrier concentrations and are now explored for use in many applications in a wide variety of fields, such as electrocatalysis and field-effect transistors.[11-14] Recently, a family of MOFs with kagomé lattices and signatures of metallicity have been reported.[15] These materials, $Ni_3(HIB)_2$ and $Cu_3(HIB)_2$, arecomposed of 2D kagomé layers in which the metal ions form vertices of the kagomé network, held together by planar HIB anions, Fig. 1.

Here we report the result of a combined thermodynamic and electrical transport study of $Ni_3(HIB)_2$ and $Cu_3(HIB)_2$. Magnetization measurements are consistent with localized S=1 and S=1/2 spin behavior, respectively, with Weiss mean-field antiferromagnetic interaction energy scales of -20.3 K and -6.52 K respectively between sites. Electrical transport reveals that both materials are small-gap semiconductors. Specific heat measurements show no evidence for a transition to a long-range ordered magnetic state above T = 0.1 K in either compound, implying a significant degree of magnetic frustration. Nonetheless, we find that structural disorder can account for all of our observations, a likely scenario since prior work has demonstrated significant stacking faults in these materials.[16]

**Experimental:**

**The synthesis procedure of $Ni_3(HIB)_2$ and $Cu_3(HIB)_2$ films (HIB = hexaiminobenzene):** $Ni_3(HIB)_2$ and $Cu_3(HIB)_2$ films were prepared using a series of synthetic procedures that are explained in detail in the literature.[1-2] Please see the summary below. *Warning: Compounds 2,4,6-trinitroaniline (TNA) and 1,3,5-triamino-2,4,6-trinitrobenzene (TATB) are very sensitive to mechanical force and highly explosive under alkaline conditions. Both should be handled with extreme caution.* Step 1: Synthesis of 2,4,6-trinitroaniline (TNA) $KNO_3$ (0.70 mol) dissolved in $H_2SO_4$ (100 mL) were added dropwise at T=50°C in a round bottom flask of 4-Nitroaniline (0.14 mol) and $H_2SO_4$. The round bottom flask was then heated to T=80°C for 3 hr and then T=110°C for 3 hr. The reaction mixture was then cooled to room temperature and later placed into an ice water container. The precipitates from the cooling were done via suction filtration followed by air drying. The recrystallized was done by 0.4 M of aqueous HCl solution to give a high yield of glass-like yellow-colored crystals of 2,4,6-trinitroaniline. Step 2: Synthesis of 1,3,5-triamino-2,4,6-trinitrobenzene (TATB). A combination of mixtures of Sodium methoxide (0.44 mol) in TNA (0.02 mol) and 4-amino-1,2,4-triazole (ATA) (0.2 mol) in DMSO (300 mL) were combined. This formed an orange-colored suspension which was stirred at room temperature for 3 hr. The reaction mixture was then poured into HCl (0.4 M), and then the precipitates separated via suction filtration, followed by washing in distilled $H_2O$ and dried. The solids were dissolved in a solution of DMSO and trace amounts of NaOH and were heated at T=70°C. After the mixture was dissolved, the mixture was decanted in a cold $HNO_3$ (0.4 M) solution. The precipitate was filtered, and deep yellow-colored TATB was obtained. Step 3: Synthesis of Hexaaminobenzene (HAB) TATB (0.012 mol), 10% Pd/C, and pure ethyl acetate were placed in a high-pressure hydrogenation bottle. The reaction container was agitated under $H_2$ (4.2 bar) for three days. The yellowish color of the reactant mixture was not observed anymore. Concentrated HCl was mixed under $H_2$ for 5 hours. The solution was filtered under reduced pressure over Celite, and HAB trihydrochloride was yielded. The precipitates were passed through suction filtration and dried at T=70 °C for 2 hr under reduced pressure. The crystals were re-dissolved in deionized $H_2O$ and filtered, concentrated aq. HCl solution was added and sealed in a freezer. This process again allowed crystallization. The crystals were accumulated through gravity filtration, washed with ethyl acetate, and dried in a furnace to yield a high yield. Step 4(a): Synthesis of $Ni_3(HIB)_2$ and $Cu_3(HIB)_2$ powders $Ni(NO_3)_2 \cdot 6H_2O$ and $CuSO_4 \cdot 5H_2O$ of 0.11 mmol were added to different scintillation vials along with 14M $NH_4OH$ in DMSO and HAB·3HCl (0.072 mmol) in degassed DMSO. The vials were heated at T=65°C for 2 hr, and a black precipitate was formed. This suspended mixture was centrifuged, decanted, and washed with deionized water and acetone. The solid product was dried under a vacuum on a Schlenk line for 10 minutes. Step 4(b): Synthesis of $Ni_3(HIB)_2$ and $Cu_3(HIB)_2$ films**.** Synthesis of $Ni_3(HIB)_2$ and $Cu_3(HIB)_2$ films were performed in a glove box. Deoxygenated ethylenediamine (50 µL) was combined with deoxygenated water (3 mL) and $Ni(NO_3)_2 \cdot 6H_2O$ (0.097 mmol) and $CuSO_4 \cdot 5H_2O$ (0.097 mmol) separately. HAB·3HCl (0.063 mmol) dissolved in deoxygenated water (3 mL) was added to the mixture. This mixture was then removed from the glovebox in a sealed container and was heated at T=40 °C for 12 hours. The MOF film appeared in a gas-liquid interface, and it was washed by deionized water and isopropyl alcohol. The crystallinity was determined using X-ray diffraction, spectroscopic analysis, and high-resolution transmission electron microscopy.

**Physical property measurement:** The resistivity and heat capacity of $Ni_3(HIB)_2$ and $Cu_3(HIB)_2$ were measured using a quantum design physical properties measurement system (PPMS). Temperature-dependent resistivity measurements were performed using a standard four-point probe technique at $\mu_oH$ = 0, 3, and 5 T. Specific-heat measurements were performed between T = 0.1–300 K at $\mu_oH$ = 0 and 1 T using the semiadiabatic pulse technique with a 1% temperature rise and measurement over three-time constants. Magnetization data were collected using the ACMS option at T = 2–200 K under $\mu_oH$ = 9T and converted to magnetic susceptibility using the approximation χ = M/H.

**Result and Discussion:**

Magnetic susceptibility measurements for $Ni_3(HIB)_2$ and $Cu_3(HIB)_2$ are shown in Fig. 2. The dilute nature of the spin centers combined with the small sample quantities necessitated the use of a large applied field, $\mu_oH$ = 9 T, in order to obtain a reliable signal. Curie-Weiss analysis on the data for Ni$_3$(HIB)$_2$ was performed for T = 150 - 200 K and yielded a Curie constant of C = 1.12 emu K mol f.u.$^{-1}$ Oe$^{-1}$ and $\Theta_{CW}$ = -20.3 K. The Curie constant is in good agreement with the expectation value of 1.0 for pure S=1 moments. Curie-Weiss analysis on the data for Cu$_3$(HIB)$_2$ was performed for T = 100 - 200 K and yielded a Curie constant of C = 0.352 emu K mol f.u.$^{-1}$ Oe$^{-1}$ and $\Theta_{CW}$ = -6.52 K. The Curie constant is in good agreement with the expectation value of 0.375 for pure S=1/2 moments. For both compounds, the Curie-Weiss behavior appears to extend down to the lowest temperature measured, T = 2 K, with no signatures of magnetic ordering. That is, they appear to exhibit magnetic frustration.

As there are many contradictory reports of metallic or semiconducting behavior in conductive MOFs, it is essential to assess the transport behavior of the materials studied here. The use of non-contact or two probe measurement configurations, combined with a high contact resistance, makes the determination of intrinsic properties difficult.[17-19] Here, a four-probe technique, which eliminates the contact resistance contributions, was utilized to re-examine the resistivity previously studied in this material at low temperatures. Both materials increase (in resistance) with decreasing temperature, Fig. 3(a) and 3(b), consistent with bulk transport dominated by a temperature-activated process. The semiconducting behavior is in agreement with a prior report that found semiconducting transport in the region T=200 – 300K; our data extends this trend to significantly lower temperatures.

An Arrhenius fit, calculated based on the relationship $\ln(\rho_T/\rho_{300\ K}) = 2k_B E_g T^{-1} + b$ where $k_B$ is the Boltzmann constant, $E_g$ is the bandgap, and $b$ is the intercept of the line, used on a linear region in a plot of the natural log of the resistivity normalized to the resistivity at $T$ = 300 K versus inverse temperature, Fig. 3(c), gives band gaps of $E_g$ = 22.2 meV for Ni$_3$(HIB)$_2$ and $E_g$ = 103 meV for Cu$_3$(HIB)$_2$.

One possible way to reconcile the contradictory reports in the literature is grain boundary effects. Grain boundary effects can result in significant changes in resistivity, including from metallic to insulating behavior, based on the compactness of the pellet (an effect well studied in PbO$_2$ pellets[25]). However, the pellets used for the resistivity measurements reported here were isobatically at 1000 bar pressed and dried with little room for grain boundaries. Although we cannot conclusively rule out grain boundary scattering as the origin of the observed temperature-activated behavior, the magnitude of the observed gaps is consistent with the observed resistances at room temperature, assuming the behavior arises from the bulk of the material and are not consistent with a composite model with metallic particles separated by an insulating matrix.[20-21]. Thus we conclude that the specimens measured here are small bandgap semiconductors. These are much smaller bandgaps than those observed in most kagomé minerals, but also far from the metallicity observed in materials such as Fe$_3$Sn$_2$ and KV$_3$Sb$_5$.[22-24]

The resistance under an applied magnetic field does not change at T=20-300K for Ni$_3$(HIB)$_2$ and T=120-300K for Cu$_3$(HIB)$_2$. However, at lower temperatures, there is a slight increase in the resistance with the application of a field. This effect most likely arises due to changes in scattering off the paramagnetic spins in these materials as they are partially polarized in the applied magnetic field.

Thermodynamic measurements provide further insight into the physical properties of M$_3$(HIB)$_2$. Fig. 4(a) shows the low temperature-specific heat of Ni$_3$(HIB)$_2$. Except for a small upturn at the lowest temperatures, we find that the data follows well the behavior expected for a simple mixture of electronic and phononic contributions: $C_p/T = \gamma + \beta_3 T^2$, where $\beta_3 T^3$ represents the phonon contribution, and $\gamma$ and the Sommerfeld coefficient. This fit gives $\beta_3$ = 4.3(8)mJ mol$^{-1}$ K$^{-4}$ and $\gamma$ = 148(4) mJ mol$^{-1}$K$^{-2}$. The latter value implies a very large and significant T-linear contribution to the specific heat – unexpected for an insulator. In the quantum spin liquid literature, such behavior is often attributed to the presence of a spinon continuum, but such behavior is prevalent in glasses without any correlated physics. Indeed, careful inspection shows that this region is not quite linear, implying a different origin, for example from a single ion singlet ground state combined with the structural disorder.[25] Further, the $\gamma$ obtained is much larger than one would expect for a typical metal (10-100 orders of magnitude larger). In addition, utilizing the $\gamma$ from heat capacity, one can estimate the carrier density, $n$, using the equation, $\gamma = 0.767\ Z\ (r_s/a_o)^2 \times 10^{-4}$ Jmole$^{-1}$K$^{-2}$, where Z = 2 and is assumed to be the ligands per unit formula for Ni$_3$(HIB)$_2$, r$_s$ is the radius of sphere per unit volume per free electrons, and $a_o$ is the Bohr radius respectively.[26-27] The carrier density, $n$ = 5.4 x $10^{19}$ cm$^{-3}$ utilizing the above case and n=1/volume and Volume= $4/3\pi r r_s^3$ relationship. This value is smaller, by a factor of more than 100, compared to the carrier density estimated using one electron per unit ligand, n = 3.9 x $10^{21}$ cm$^{-3}$. Both estimates have their limits, as carrier density assumed from heat capacity assumes a simple Fermi surface and no other T-linear contributions such as from glassy behavior, and the second case considers one electron per unit volume. Nevertheless, these values suggest that $Ni_3(HIB)_2$ is a semiconductor. The small upturn starting below T ~ 5 K develops into a broad hump below T ~ 1 K, Fig. 4(b). It is likely not an intrinsic magnetic contribution, as the change in entropy is very small (inset Fig.4(a)), and is more likely the result of defect spins or other disorders in the material.

The low-temperature specific heat of $Cu_3(HIB)_2$ does not follow simple relations under any measured conditions. The low-temperature heat capacity of $Cu_3(HIB)_2$, Fig. 4(b), shows a smooth divergence down to T = 0.1 K. This divergence moves to higher temperatures and becomes more pronounced under the application of a $\mu_o H$ = 1 T field, suggesting a magnetic origin. Further, the magnitude of the change is significant, implying that the behavior at these temperatures is not governed by strong interactions between spins (which would tend to suppress the ability of an applied field to change the magnetic behavior). Nonetheless, no signatures of a phase transition are observed.

In light of these results, it is important to ask what the likely origins of our observations are. Lack of magnetic order or other phase transitions and T-linear contributions to specific heat is often taken as quantum spin liquid physics signatures. However, substantial work in recent years has shown how disorder can often present many of the same attributes without the presence of a QSL state and also how disorder might itself enable QSL-type states. [28-32]

For the present materials, prior work reported synchrotron X-ray diffraction data that indicated a substantial number of stacking faults between distinct layers[5]. These stacking faults cause a characteristic signature of structural disorder, which, as in the case of $MoS_2$ or graphite, most likely arises due to the non-uniformity of the stacking of well-defined 2D layers.[33-36] Thus, we attribute the observed behaviors not to quantum spin liquid physics but rather to the disorder-driven spreading of states in energy and momentum, avoiding the well-defined features necessary for order to arise.

**Conclusion:**

Overall, the magnetization, heat capacity, and resistivity data of $Ni_3(HIB)_2$ and $Cu_3(HIB)_2$ suggest that these materials are paramagnets with antiferromagnetic correlations and small, semiconducting bands gaps. Although we do not observe a sizeable effect in the transport measurements on applying a magnetic field, we observe a 2.5-fold increase in the heat capacity for $Cu_3(HIB)_2$ with $\mu_o H$ = 1 T. This trend indicates nearly free spins easily perturbed by an external field and rules out strongly interacting spin liquid physics in this regime. Similarly, for $Ni_3(HIB)_2$, the large T-linear specific heat can be explained by a confluence of a single ion singlet ground state with stacking faults due to the disorder between layers. Our results further state the art of kagomé lattice physics, especially in the rarely explored regime of semiconducting but not metallic behavior.

**Acknowledgments:**

This work was supported as part of the Institute for Quantum Matter, an Energy Frontier Research Center funded by the United States Department of Energy, Office of Science, Office of Basic Energy Sciences, under Award DE-SC0019331. KEA acknowledges the funding provided by the Alexander von Humboldt Foundation. Work in the Dincă lab was funded by the Army Research Office (Award No. W911NF-17-1-0174).

**Conflicts of interest:**

There are no conflicts to declare.

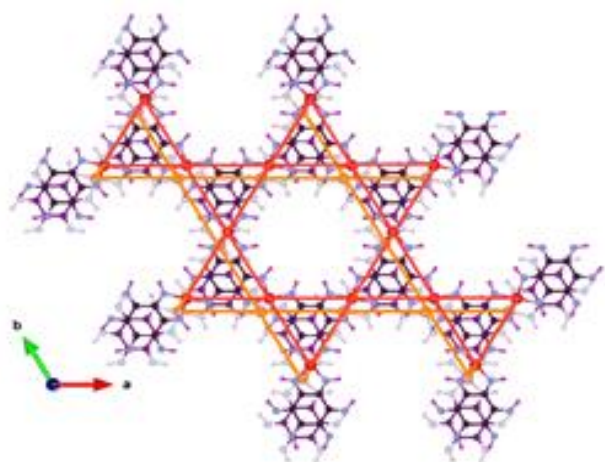

**Figure-1** Kagomé network of M₃(HIB)₂ denoted in red and orange lines amongst the metal ion denoted in red circles.

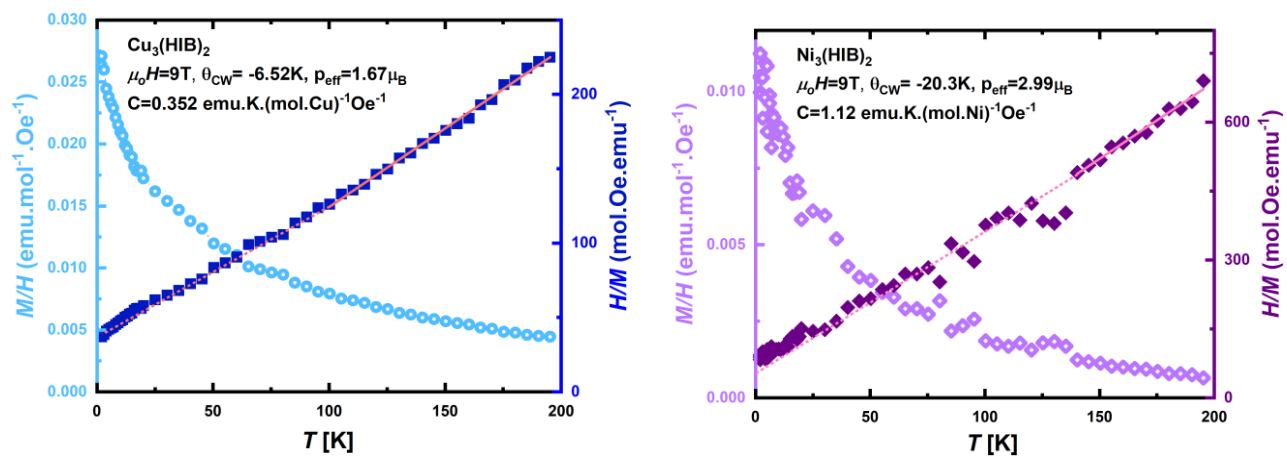

**Figure-2** Magnetic susceptibility, estimated as M/H, and Curie-Weiss analysis, of $Cu_3(HIB)_2$ and $Ni_3(HIB)_2$. The data suggests S=1/2 and S=1 paramagnetic behavior, respectively, with weak antiferromagnetic interactions. This data was collected at $\mu_o H$ = 9T.

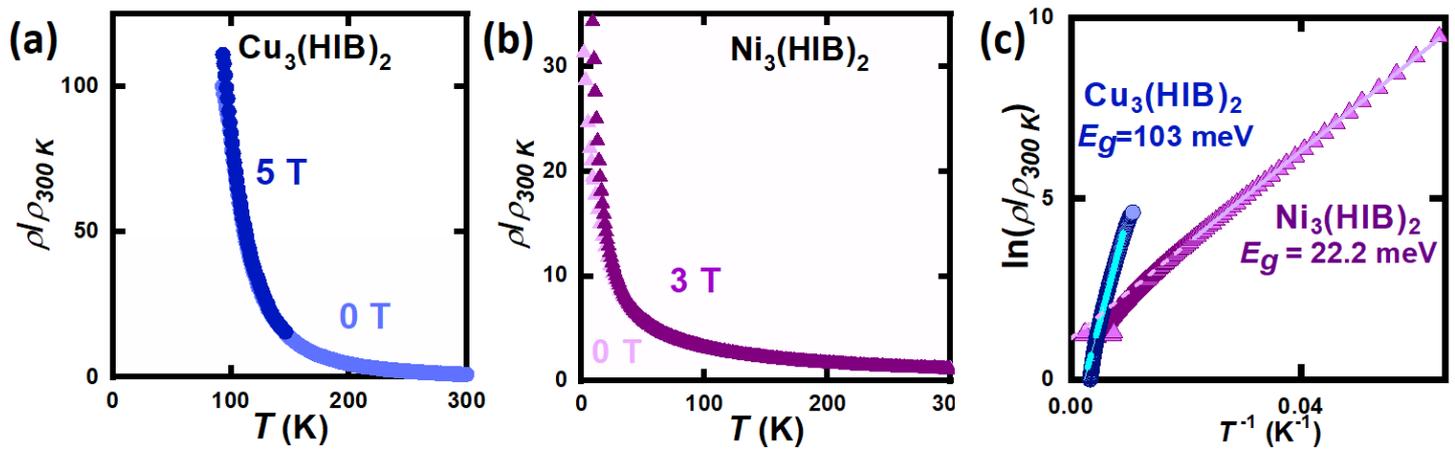

**Figure-3(a)** Resistivity of $Cu_3(HIB)_2$ at $\mu_oH$ = 0 and 5 T. **(b)** Resistivity of $Ni_3(HIB)_2$ at $\mu_oH$ = 0 T and 3 T. **(c)** Arrhenius fit for bandgap determination of $Cu_3(HIB)_2$ is $E_g$ = 103 meV and $Ni_3(HIB)_2$ is $E_g$ = 22.2 meV.

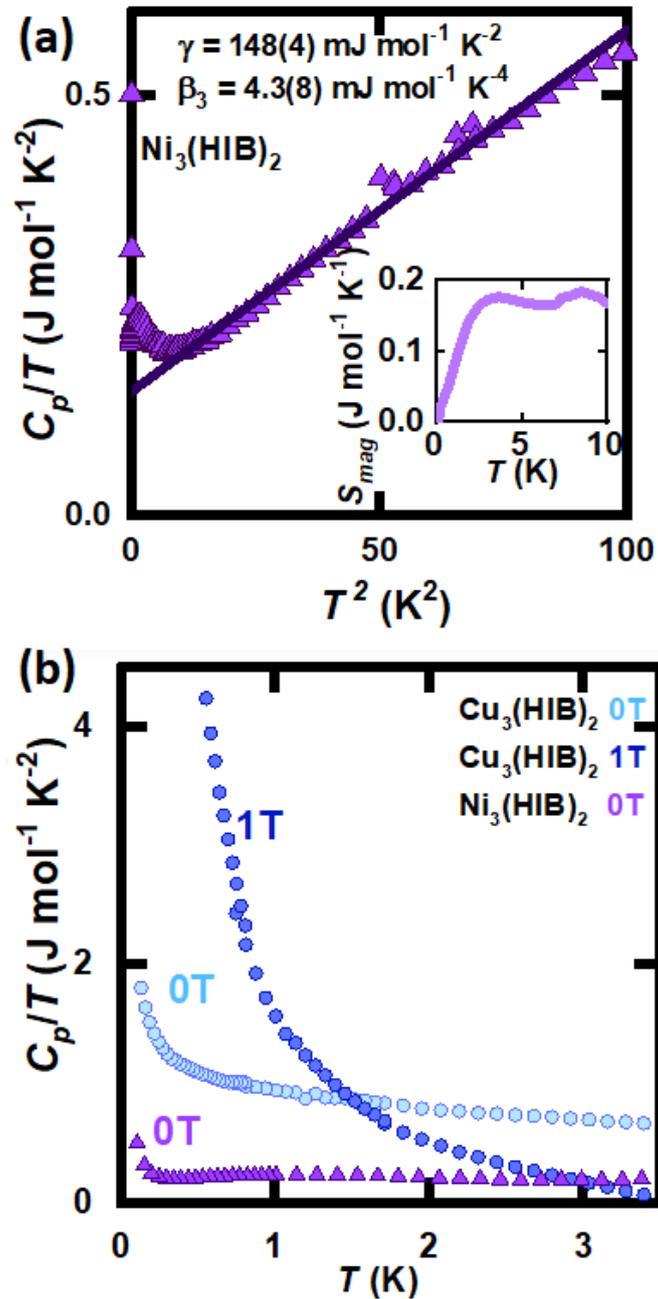

**Figure-4 (a)** Linear fit of heat capacity divided by temperature, $C_p/T$ as a function of $T^2$ plot of Ni$_3$(HIB)$_2$ where $C_p/T = \gamma + \beta_3 T^2$, where $\beta_3 T^3$ represents the phonon contribution, and $\gamma$ and the Sommerfeld coefficient. The inset is the change in excess entropy from T=0 to 10 K. **(b)** $C_p/T$ as a function of T plot where Cu$_3$(HIB)$_2$ data for *at $\mu_o H$ = 0T and 1T* is compared to the Ni$_3$(HIB)$_2$ $\mu_o H$ = 0T.